\documentclass[submission, Proceedings]{SciPost}
\usepackage{lineno}
\DeclareUnicodeCharacter{2212}{-}

\hypersetup{
    colorlinks,
    linkcolor={red!50!black},
    citecolor={blue!50!black},
    urlcolor={blue!80!black}
}

\usepackage[bitstream-charter]{mathdesign}
\DeclareSymbolFont{usualmathcal}{OMS}{cmsy}{m}{n}
\DeclareSymbolFontAlphabet{\mathcal}{usualmathcal}

\begin{document}

\begin{center}{\Large \textbf{B boson search at KLOE/KLOE-2}}\end{center}

\begin{center}
Elena P\'erez del R\'io \textsuperscript{1$\star$} for the KLOE-2 Collaboration
\end{center}

\begin{center}
{\bf 1} Jagiellonian University, Krak\'ow, Poland
\\
* elena.rio@uj.edu.pl
\end{center}

\begin{center}
\end{center}


\definecolor{palegray}{gray}{0.95}
\begin{center}
\colorbox{palegray}{
  \begin{minipage}{0.95\textwidth}
    \begin{center}
    {\it  16th International Workshop on Tau Lepton Physics (TAU2021),}\\
    {\it September 27 – October 1, 2021} \\
    \doi{10.21468/SciPostPhysProc.?}\\
    \end{center}
  \end{minipage}
}
\end{center}

\section*{Abstract}
{\bf
Following the many contributions KLOE-2 has done to Dark Matter (DM) searches, an alternative model, where the Dark Force mediator is an hypothetical leptophobic B boson, in contra-position to the U boson or "dark photon", is investigated. The B boson couples mainly to quarks and it can be searched in the Phi decay to $\eta$-B where B will decay in $\pi^{0}$-$\gamma$. So far, investigation of the $\pi^{0}$-$\gamma$ invariant mass shows no clear structure belonging to the signal of the DM mediator, hence, an upper limit in the number of events at 90\% with CLs the technique will be established for the decay.
}


\section{Introduction}
\label{sec:intro}
The Universe consists, in its majority, of Dark Matter (DM) and Dark Energy, being the contribution of visible matter as little as 4\% of the total.
While the existence of DM is well established nowadays and supported by cosmological observations, the nature of the same is still an unknown. Many models try, from different approaches, to explain he elusive matter. Some of this attempts open the possibility of new hidden forces not included in the Standard Model (SM) of particles, opening a way for New Physics. One of the direct implications of such a new matter would serve as an explanation for the g-2 anomaly~\cite{AOYAMA20201}, the longest standing discrepancy between theory and experimental values in the SM, which has been confirmed at the level of 4.2$\sigma$~\cite{PhysRevLett.126.141801}. Several extensions of the SM have proposed models~\cite{ext1, ext2, ext3, ext4,ext5} with a Weakly Interacting Massive Particle
(WIMP) belonging to a secluded gauge sector. The minimal case describes the new gauge interaction mediated by a new vector gauge boson, the U boson or {\it dark photon}, which could interact with the photon via a kinetic-mixing term. 
After several years of sterile searches of the U boson, specially in the $g-2$ preferred region, new models are gaining popularity, like the existence of a leptophobic {\it dark photon}~\cite{sean}, a new gauge field coupling to baryon number. This new dark matter mediator, or B boson, would mainly decay to baryons, opening new possibilities of discovery.

\section{KLOE at DA\texorpdfstring{$\Phi$}{Lg}NE}
\label{sec:kloesetup}
The KLOE (K LOng Experiment) and KLOE-2 experiments, at the DA$\Phi$NE $\phi$-factory at the Laboratori Nazionali di Frascati in Italy, were able to measure mesons in the range of energies from 5 MeV to 1 GeV. The experimental setup of KLOE consisted of a large cylindrical drift chamber (DC)~\cite{DC} surrounded by an electromagnetic calorimeter made of lead-scintillating fibers~\cite{EMC}, all embedded in a magnetic field provided by a superconductive coil. The energy and time capabilities of the EMC, with resolutions of $\sigma_E/E = 5.7\%/\sqrt{E\text{[GeV]}}$ and $\sigma_t(E)=57\text{ps}/\sqrt{E\text{[GeV]}}\oplus100 \text{ps}$, respectively, and the DC position resolutions of $\sigma_{xy} \sim 150 \mu\text{m}$ and $\sigma_z \sim 2\text{mm}$ and momentum resolution, $\sigma_{p\perp}/p_{\perp}$, better than $0.4\%$ for large angle tracks, made of KLOE a perfect setup to study kaon-, and hadron-physics, and rare meson decays.

The KLOE experiment collected $2.5~\mathrm{fb^{-1}}$ at the $\phi$-peak during its 2002 to 2005 campaign and some more data off-peak. KLOE-2, the upgraded continuation of the experiment, acquired data from late 2014 to early 2018, collecting more than $5~\mathrm{fb^{-1}}$, thanks to an upgraded beam crossing scheme of the DA$\Phi$NE collider.
For the KLOE-2 run the original detector setup was extended with the installation of a inner tracker~\cite{cgem-it} and two calorimeters~\cite{CCALT,QCALT} close to the interaction region (IP), in order to improve the vertex reconstruction near the interaction point (IP) and increase tightness of the detector. Additionally, two couples of energy taggers~\cite{HET-LET} were installed to study $\gamma\gamma$ fusion. The full sample of KLOE and KLOE-2 data presents an invaluable and unique collection to carry out a broad program involving, kaon-, hadron-physics and dark matter searches among others~\cite{KLOE2_proposal,doi:10.1142/S0217751X19300126}.

\section{Leptophobic B boson}
\label{sec:another}
As mentioned in the introduction~\ref{sec:intro}, WIMPS have been proposed as plausible components of the DM component of our Universe. The interaction between these candidates and the SM particles is commonly proposed to be mediated by a new vector gauge boson called {\it{dark photon}}, U or A', which couples to the photon via a kinematic mixing term, $\epsilon^2$, which can be searched for in $e^+e^-$ colliders via processes  $e^+e^- \rightarrow U \gamma$, $V \rightarrow P\gamma$ decays, where V and P are vector and pseudoscalar mesons, and $e^+e^- \rightarrow h'U$, where $h'$ is a Higgs-like particle responsible for the breaking of the hidden symmetry. The KLOE collaboration has profusely contributed to these searches studying different processes~\cite{combined_limit,enrico, KLOE_UL1, KLOE_UL2, ppg}, covering a large part of the expected mass region. 

\begin{figure}
    \centering
    \includegraphics[width=10cm]{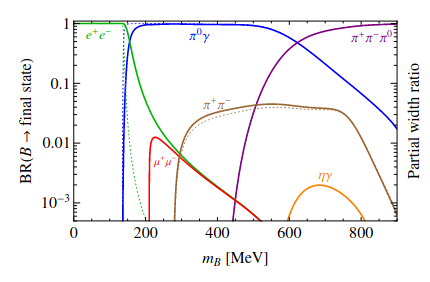}
    \caption{Figure from~\cite{sean}. branching ratios for B decay. Thick lines have $\epsilon = eg_B/(4\pi)2$; thin dotted lines have $\epsilon = 0.1 \times eg_B/(4\pi)2$.}
    \label{fig:brBboson}
\end{figure}

Additionally, a new model where the interaction is mediated by new gauge boson that couples preferably to quarks over leptons, as proposed in the U photon models, thus being quoted as leptophobic. In the simplest model, the new baryonic-force mediator, called B boson, would couple to baryon number and arise from a new $U(1)_{B}$ gauge symmetry~\cite{sean}. In this case, the interaction the lagrangian is described by:
\begin{equation}
    \mathcal{L} = \frac{1}{3}g_{B}\Bar{q}\gamma^{\mu}qB_{\mu}
    \label{eq:lagrangian}
\end{equation}
where $B_{\mu}$ is the new gauge field coupling to baryon number. The gauge coupling $g_{B}$ is universal for all quarks q. From this, a baryonic fine structure constant $\alpha_B  \equiv g^2_B/(4\pi)$, analogous to the electromagnetic constant $\alpha_{em}$, is defined.
Since eq.~\ref{eq:lagrangian} preserves all low-energy symmetries of QCD and the gauge coupling is universal for all flavors, B does not transform under the flavor symmetry and is a singlet under isospin. This means that B has assigned the same quantum numbers as the $\omega$ meson: $I^G(J^{PC}) = 0^-(1^{--})$. Thus, representing a good guide of what we can expect for the decays of the B boson. From $\omega$ we know that the three main decay modes Branching Fractions are: $BR(\omega \rightarrow \pi^+ \pi^- \pi^0) \simeq 89\%$, $BR(\omega \rightarrow \pi^0 \gamma) \simeq 8\%$ and $BR(\omega \rightarrow \pi^+ \pi^- \simeq 1.5\%$. This scheme is expected be followed by the B boson in the range $m_{\pi} \le  m_B \le \text{GeV}$. The different branching fractions can be seen in the fig.~\ref{fig:brBboson}. Above 1 GeV, the kaon channel $B \rightarrow K \bar{K}$ opens and the B decays become similar to those of the $\Phi$ meson.

\subsection{Search of the B boson with KLOE detector}
As seen in fig.~\ref{fig:brBboson}, in the mass range between $m_{\pi}$ and $m_{3\pi} \approx 620 \text{MeV}$, the dominant channel is the decay $B \rightarrow \pi^0 \gamma$. With this in mind, we perform the search of the B boson with the KLOE data by looking for an enhancement in the invariant mass of $m_{\pi^0\gamma}$ from the decay chain $\Phi \rightarrow B \eta \rightarrow \pi^0 \gamma$, with $eta \rightarrow \gamma\gamma$ and $\pi^0 \rightarrow \gamma\gamma$.

A data sample of $1.7 fb^{-1}$ has been analyzed looking for $5-\gamma$ final states. For this, exactly five neutral clusters with a deposited energy in the calorimeter above $10 \text{MeV}$ and scattering angle $23\circ \le \theta \le 157\circ$, this is in the barrel section of the KLOE calorimeter, are selected. To this selection, a kinematic fit is applied to improve the resolution and allow to properly select the $\eta-\pi^0$ containing events. 

\begin{figure}
    \centering
    \includegraphics[width=8cm]{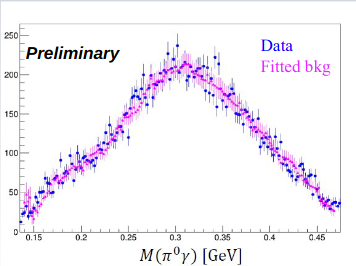}
    \caption{Preliminary spectra of the invariant mass of the $\pi^0\gamma$ system. The B boson signature would be a pronounce enhancement in the spectra. Full blue circles represent the data and full magenta circles the extrapolated background using a side-band fit.}
    \label{fig:Bmass}
\end{figure}

After a refine selection and proper background discrimination, the mass of the $\pi0\gamma$ system can be reconstructed, as presented in fig.~\ref{fig:Bmass} in blue full circles. The expected background according to Monte Carlo simulations corresponds to the irreducible SM backgrounds coming from the decays $\Phi \rightarrow a_0 \gamma \rightarrow \eta \pi^0 \gamma$ and $\Phi \rightarrow \eta \gamma$ with $\eta \rightarrow 3\pi^0$, where final state photons can be undetected, therefore lost, or merged to another hits by the reconstruction algorithms. To avoid model dependency, the background in the invariant mass of $\pi^0\gamma$ is extracted by a side-band fit, and the interpolated background with this method is shown in fig.~\ref{fig:Bmass} as magenta full circles.

\begin{figure}
    \centering
    \includegraphics[width=12cm]{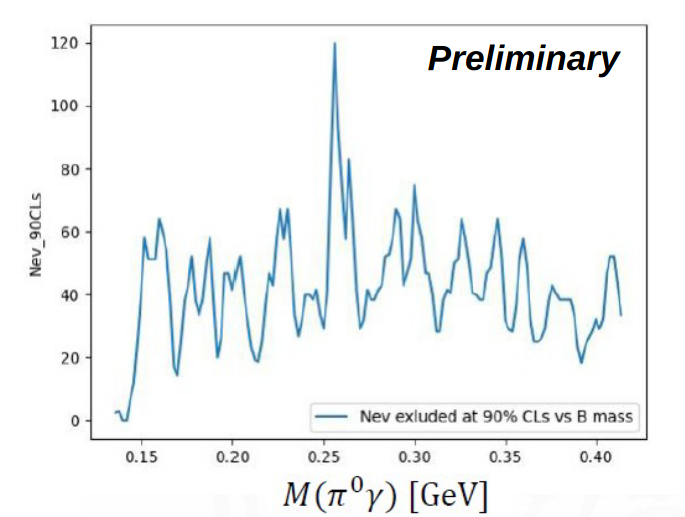}
    \caption{Preliminary Upper Limit at 90\% CLs in the number of signal events using CLs method.}
    \label{fig:Nsig}
\end{figure}

Since no excess is observe in the spectra, we proceed to extract the upper limit in the B boson by using the CLs method~\cite{cls}. The preliminary result of the upper limit on the number of expected signal events at 90\% CLs is presented in fig.~\ref{fig:Nsig}. From this we can expect to set limits in the coupling constant of the B boson ($\alpha_B$) at the level of $\mathcal{O}(10^{-7})$ at 90\% CLs.

\section{Conclusion}
The KLOE-2 Collaboration continues contributing to the searches of DM and Physics Beyond the Standard Model by exploring the reaction $\Phi \rightarrow B \eta$ with B being a new gauge boson coupling mainly to quarks. The interested decay channel of this study is the process $B \rightarrow \pi^0 \gamma$, which can be mimicked by the SM process $\Phi \rightarrow a_0 \gamma$ and $\Phi \rightarrow \eta \gamma \rightarrow 3\pi^0 \gamma$ when photons are not detected or merged to another hits during the reconstruction. In the scenario of the discovery of the new mediator, the evidence would be an enhancement in the invariant mass of the $\pi^0\gamma$ pairs. Since no evidence of the B boson is found, we proceed to extract the upper limit on the coupling of the B boson to the SM particles, $\alpha_B$. In this work we present the preliminary upper limit in the number of B boson signal events extracted with the CLs method at 90\%, from which we expect to set limits of the oder of $\mathcal{O}(10^{-7})$ at 90\% CLs in the coupling constant.

\section*{Acknowledgements} 
We warmly thank our former KLOE colleagues for the access to the data collected during the KLOE  data-taking campaign. We thank the DA$\Phi$NE team for their efforts in maintaining low background running conditions and their collaboration during all data taking. We want to thank our technical staff: G.F. Fortugno and F. Sborzacchi for their dedication in ensuring efficient operation of the KLOE computing facilities; M. Anelli for his continuous attention to the gas system and detector safety; A. Balla, M. Gatta, G. Corradi and G. Papalino for electronics maintenance; C. Piscitelli for his help during major maintenance periods. This work was supported in part by the Polish National Science Centre through the Grants No. 
2013/11/B/ST2/04245, 2014/14/E/ST2/00262, 2014/12/S/ST2/00459,
\\
2016/21/N/ST2/01727,2016/23/N/ST2/01293, 2017/26/M/ST2/00697.





\bibliography{bibliography.bib}

\nolinenumbers

\end{document}